# DAΦNE GAMMA-RAYS FACTORY

*D. Alesini[1], I. Chaikovska[2], A. Variola[2], S. Guiducci[1], F. Zomer[2], C. Milardi[1] and M. Zobov[1]*

[1]LNF-INFN, Via E. Fermi 40, 00044 Frascati, Rome, Italy.

[2]Laboratoire de l'Accélérateur Linéaire, CNRS-IN2P3, Université Paris-Sud 11, 91898 Orsay, France

## Abstract

Gamma sources with high flux and spectral densities are the main requirements for new nuclear physics experiments to be performed in several worldwide laboratories and envisaged in the ELI-NP (Extreme Light Infrastructure-Nuclear Physics) project or in the IRIDE (Interdisciplinary Research Infrastructure with Dual Electron Linacs) proposals. The paper is focalized on an experiment of gamma photons production using Compton collisions between the DAFNE electron beam and a high average power laser pulse, amplified in a Fabry-Pérot optical resonator. The calculations show that the resulting gamma beam source has extremely interesting properties in terms of spectral density, energy spread and gamma flux comparable (and even better) with the last generation gamma sources. The energy of the gamma beam depends on the adopted laser wavelength and can be tuned changing the energy of the electron ring. In particular we have analyzed the case of a gamma factory tunable in the 2-9 MeV range. The main parameters of this new facility are presented and the perturbation on the transverse and longitudinal electron beam dynamics is discussed. A preliminary accelerator layout to allow experiments with the gamma beam is presented with a first design of the accelerator optics.

# 1    Introduction

Gamma-ray sources with high flux and high spectral densities can open new frontiers in the nuclear physics and in the nuclear photonics experiments with many fundamental studies in nucleus structure and strategic applications for nuclear waste treatment, national security, nuclear medicine, astrophysics and nucleosynthesis. Many projects worldwide, based on Compton back-scattering between electron bunches and counter propagating laser pulses, are going in this direction and several laboratories are pursuing projects to develop such advanced sources. The main ones are Mega-ray project [1], AIST [2], and ELI-NP [3,4] with the aim to overcome by few orders of magnitude the present state of the art parameters of the facility HIGS [5] in operation since few years with mono-chromatic gamma ray beams. In particular, at the Nuclear Physics Pillar of the European laser facility ELI (Extreme Light Infrastructure), to be developed in Romania at Magurele (near Bucharest), an advanced Gamma System is foreseen as a major component of the infrastructure, aiming at producing extreme gamma ray beams for nuclear physics and nuclear photonics experiments for users. The challenging requirements of this source are the tunable energy of the gamma beam (1–20 MeV), the very narrow bandwidth (0.3%), and the high spectral density ($10^4$ photons/s/eV). This source is based on a 100 Hz repetition rate RF Linac operated at C-band [6] with an S-band photo-injector delivering a high phase space density electron beam in the 180-720 MeV energy range colliding with a Yb:Yag high power laser pulse, to produce via Compton back-scattering a gamma-ray photon beam. In order to increase the average number of photons a recirculation of the high intensity interaction laser is needed [3] leading to an effective rep. rate of the collision of about 3 kHz.

Also the project Mega-ray is based on a multi-bunch X-band linac colliding with a laser [1].

Other gamma sources [2,8,9], indeed, produce Compton back-scattering photons in MeV range, with an electron storage ring and high-power laser systems [2,3]. Some of them use free-electron lasers instead of the conventional ones to increase photon yield [5, 10–12].

To obtain high flux and, at the same time, high spectral densities gamma sources it is necessary both to increase the number of electron-photon collisions and to control the electron and laser beam qualities. As explained in [13], in normal conducting photo-injectors the maximum rep. rate of the collisions cannot overcome few kHz and also the bunch charge cannot overcome few hundred of pC to preserve good beam quality. On the other hand, thanks to the very good beam emittance, it is possible to strongly focalize the beam at the interaction point (IP) thus increasing the number of the emitted gamma photons without affecting the resulting gamma beam bandwidth. Colliding lasers with a high energy per pulse and low rep. rate are then necessary. As example in the European ELI-NP proposal [3-4] a 100 Hz laser has been considered with 0.5 J laser pulses and a recirculation scheme of the laser to increase the collision repetition rate [4].

High flux and high spectral densities can be also obtained using storage ring and high rep. rate laser. The high flux can be obtained increasing the current per bunch and the energy of the laser per pulse while the high spectral density can be obtained with a proper choice of the machine and laser parameters.

The aim of the proposal presented in this paper is to create such a source using the DAFNE stored electron beam colliding with a laser beam amplified in a Fabry-Perot cavity (FPC). DAFNE is an $e^+e^-$ collider operating at the energy of Φ-resonance (1.02 GeV c.m.) [18]. The main machine parameters [19-20], are summarized in Table I. As discussed in the paper, the extremely high current storable in the electron ring (>2 A) and the achievable beam emittance, allow reaching excellent gamma beam qualities (in terms of both flux and spectral densities) comparable (and even better) with those of the new generation gamma sources. As discussed in the text, with a proper choice of the machine parameters, the interval between two consecutive collisions of an electron with a photon can be much longer than the damping time of the machine and the beam dynamics is completely dominated by the dynamics of the electron ring without the laser. A first test of collision between a stored electron beam and a laser beam stored in a FPC to generate gamma photons has been done recently at ATF [22]. What we propose in DAFNE has a similar scheme but the resulting gamma source flux and the beam quality are orders of magnitude higher than those achievable in the present facilities. The FPC that we propose to use in DAFNE is similar to the one designed and fabricated at LAL Orsay, and used in the ATF experiment [23]. This cavity is expected to have similar wavelength, repetition rate, stored power and beam focalization at the Interaction Point (IP). The energy of the gamma source can be changed by changing the energy of the electron ring in the range 2-9 MeV. Longer laser wavelength can be in principle used allowing to explore different ranges of gamma beam energies up to 200 keV even if, in this case, due to the available technology, lower laser power will be available and, as a consequence, lower gamma flux can be achieved.

In the first section of the paper we briefly review the basic formulas that characterize Compton-based gamma sources. In the second section we present the results of the calculations related to the DAFNE-based gamma source. In the third section we discuss the perturbation in the electron beam dynamics taking into account the beam-laser interaction. Simulations results with the Monte Carlo code CAIN [25] are given in the fourth section. In the fifth section a preliminary schematic layout of the gamma source is presented with few considerations on a possible use of the FPC LAL cavity. Results of the DAFNE low emittance optics studies are given in the sixth section.

Table I: DAΦNE parameters

| | | |
|---|---|---|
| Energy | E [MeV] | 510 |
| Machine length | L [m] | 97.6 |
| Max. achieved e⁻ beam current | $I_{MAX}$ [A] | 2.5 (in 115 bunches over 120) |
| RF frequency | $f_{RF}$ [MHz] | 368.67 |
| Max RF voltage | $V_{RF\_MAX}$ [kV] | 250 |
| Harmonic number | $h_N$ | 120 |
| Minimum bunch spacing | $T_B$ [ns] | 2.7 (=1/$f_{RF}$) |
| Horizontal emittance | $\varepsilon_x$ [mm mrad] | 0.250 |
| Coupling | $coupl=\varepsilon_y/\varepsilon_x$ [%] | <0.5 |
| Bunch length | $\sigma_t$ [ps] | 40 (low current)-60 (high current) |
| Energy spread | $\Delta\gamma/\gamma$ [%] | 0.04 (low current)-0.06 (high current) |
| Longitudinal damping time | $\tau_{damp}$ [ms] | 17 |

## 2     Basic formulas for gamma source characterization

Compton sources can be considered as electron-photon colliders. For a generic source there are four important quantities that characterize the source: the total number of scattered photons per second over the 4π solid angle, the rms source bandwidth, the number of photons per second in the bandwidth and the spectral density. These quantities can be calculated with simple formulas that can be found in [4,14-18] and that we report in the following for completeness.

In the case of gaussian beams, the total number of scattered photons per second over the 4π solid angle can be calculated by the formula given in [22, 32]:

$$N_\gamma = \sigma_{Compton} \cdot L = \sigma_{Compton} \frac{N_e N_{ph} f_{coll} \cos(\phi/2)}{2\pi\sqrt{\sigma_{y\_e}^2 + \sigma_{y\_L}^2}\sqrt{(\sigma_{x\_e}^2 + \sigma_{x\_L}^2)\cos^2(\phi/2) + (\sigma_{z\_e}^2 + \sigma_{z\_L}^2)\sin^2(\phi/2)}} =$$

$$= 4.2 \cdot 10^8 \frac{U_L[J]Q[pC]f_{coll}\cos(\phi/2)}{h\nu_L[eV]\sqrt{\sigma_{y\_e}^2 + \sigma_{y\_L}^2}\sqrt{(\sigma_{x\_e}^2 + \sigma_{x\_L}^2)\cos^2(\phi/2) + (\sigma_{z\_e}^2 + \sigma_{z\_L}^2)\sin^2(\phi/2)}[\mu m^2]}$$

(1)

where $\sigma_{Compton}$ is the Compton cross section, $N_e$ and $N_{ph}$ are the number of electrons per bunch and the number of photons in the laser pulse, $\phi$ is the collision angle, $f_{coll}$ is the collision frequency, $\sigma_{x,y,z\_e}$ and $\sigma_{x,y,z\_L}$ are the rms electron bunch and laser pulse sizes at the IP, $U_L$ and Q are the laser pulse energy and the electron bunch charge, $h\nu_L$ is the laser photon energy.

The wavelength of the emitted radiation [15-16] within a small angle of scattering with respect to the propagation axes of the electron beam, is related to the laser wavelength $\lambda_L$ by:

$$\lambda_\gamma \cong \lambda_L \frac{1+\gamma^2\theta^2+\phi^2/4}{4\gamma^2}$$

(2)

where $\theta$ is the observer angle and $\gamma$ the electron relativistic factor.

The rms bandwidth of the emitted photons can be calculated by the following formula [3,4, 14-17]:

$$\frac{\Delta\nu_\gamma}{\nu_\gamma} \cong \sqrt{(\gamma\vartheta)^4 + 4\left(\frac{\Delta\gamma}{\gamma}\right)^2 + \left(\frac{\varepsilon_n}{\sigma_x}\right)^4 + \left(\frac{\Delta\nu}{\nu}\right)^2 + \left(\frac{M^2\lambda_L}{2\pi w_0}\right)^4 + \left(\frac{a_{0p}^2/3}{1+a_{0p}^2/2}\right)^2}$$

(3)

where $\gamma\vartheta$ is the normalized collecting angle; $\Delta\gamma/\gamma$ is the electron bunch energy spread, $\varepsilon_n$ is the normalized transverse emittance (assuming round beams), $\Delta\nu_L/\nu_L$ is the relative laser pulse bandwidth, $a_{0p} = 4.3\frac{\lambda_L}{w_0}\sqrt{\frac{U[J]}{\sigma_{z\_L}/c[ps]}}$ is the laser parameter, $w_0$ is the laser waist ($w_0=2\sigma_{x,y\_L}$) and $\frac{M^2\lambda_L}{2\pi w_0}$ is the laser diffraction broadening contribution. The contribution of the normalized collecting angle and beam energy spread is a direct consequence of the dependence of the emitted radiation from the observer angle and beam energy as given in eq. (2). The contribution $\varepsilon_n/\sigma_x$ is due to the fact that, since the electron beam has a divergence at the IP, different electrons collide with photons at different angles and this contributes to the gamma energy spread according to eq. (2).

The number of photons per second in the bandwidth for small normalized collecting angles, $(\gamma\vartheta)^2 \ll 1$, can be calculated approximating the formulas given in [16,21] (see appendix 1):

$$N_\gamma^{bw} \cong 6.2\cdot 10^8 \frac{U_L[J]Q[pC]f_{coll}\cos(\phi/2)}{h\nu_L[eV]\sqrt{\sigma_{y\_e}^2+\sigma_{y\_L}^2}\sqrt{(\sigma_{x\_e}^2+\sigma_{x\_L}^2)\cos^2(\phi/2)+(\sigma_{z\_e}^2+\sigma_{z\_L}^2)\sin^2(\phi/2)}[\mu m^2]}(\gamma\vartheta)^2$$

(4)

The spectral density, defined by the ratio between the number of photons in the bandwidth and the rms value of the bandwidth, can be finally calculated by:

$$SPD\left[\frac{ph}{s \cdot eV}\right] \equiv \frac{N_\gamma^{bw}}{\sqrt{2\pi}h\Delta\nu_\gamma}$$

(5)

## 3    Collision scheme with flat electron beam and round laser beam

A simple sketch of a possible design of the interaction region is given in Fig. 1 while two possible collision schemes are given in Fig. 2. The electron and the laser bunch lengths correspond to typical values of DAFNE and FPC bunch lengths [23] and the two beams collide with an angle ϕ=8 deg (equal to the angle already implemented in the ATF experiment [22]).

In the first case a full coupling in the electron beam is assumed, while, in the second case, a small value of the coupling is considered. In this last configuration, which is the normal configuration for the DAFNE collider, where the two oscillations planes are decoupled, we have, in principle, two different values of the parameter $\varepsilon_n/\sigma_{x,y}$ and, therefore, a different contribution of the beam divergence for the gamma energy spread. To have an equal contribution in both planes it is easy to verify that the ratio of the β-functions at the IP $\beta_y/\beta_x$ has to be equal to the machine coupling.

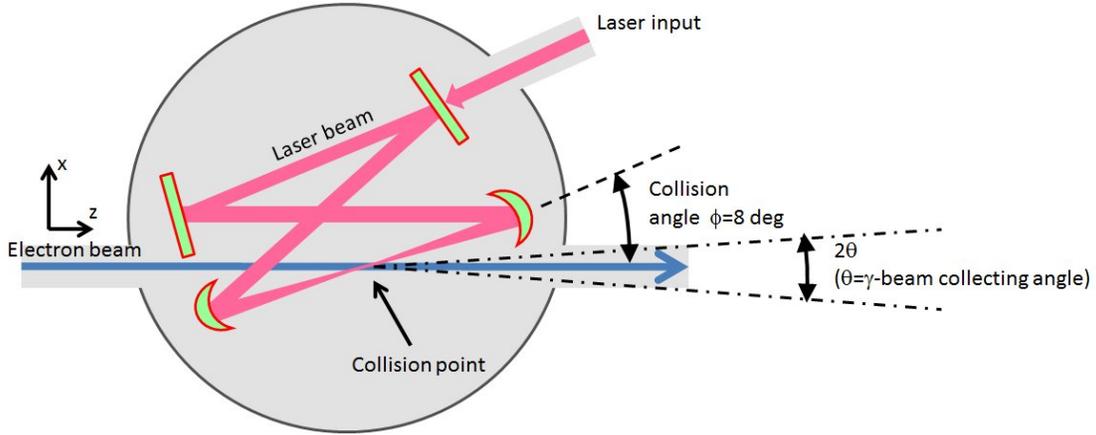

Fig. 1: simple sketch of the interaction region

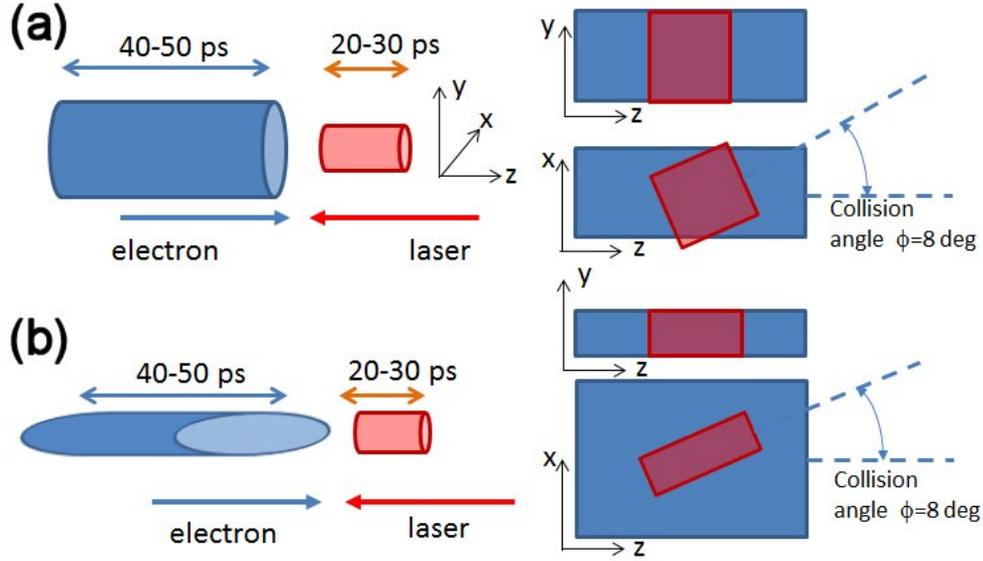

Fig. 2: two possible collision schemes.

We have analyzed the case of a flat electron beam and a round laser beam with equal vertical dimensions at the IP. The results are given in Fig. 3-5 where the gamma source parameters have been plotted as a function of the laser spot size. For the rms bandwidth calculation we have assumed that the main contribution to the energy spread is due to the collection angle and e⁻ beam parameters. Furthermore we fixed the collecting angle in order to have a contribution in the energy spread coming from the collecting angle equal to the other terms ( $\vartheta = (1/\gamma)\left[4(\Delta\gamma/\gamma)^2 + (\varepsilon_n/\sigma_x)^4\right]^{1/4}$ ). For DAFNE we have considered the case of 1.5 A in 60 equally spaced bunches that is compatible with the best performances of the machine in the last SIDDARTHA run. The emittance has been considered to be 40% its present value. As shown in section 7 this value can be achieved in DAFNE with a proper optics design. Concerning the coupling values of 2.5% and 1%, this is perfectly compatible with the machine operation since we normally reach values lower than 1%. For the laser we have considered the case of $\lambda_L=1\mu m$ with a collision angle between electron and laser of 8 degrees, the collisions repetition frequency is 184 MHz ($f_{RF}/2$). All these parameters are compatible with those of the FPC used in ATF. The choice of the wavelength fixes the maximum energy of the gamma beam according to eq. (2).

The results show the extremely good quality of the source both in term of flux and in term of SPD. The best results have been summarized in Table II for two cases of 1% and 2.5% of coupling.

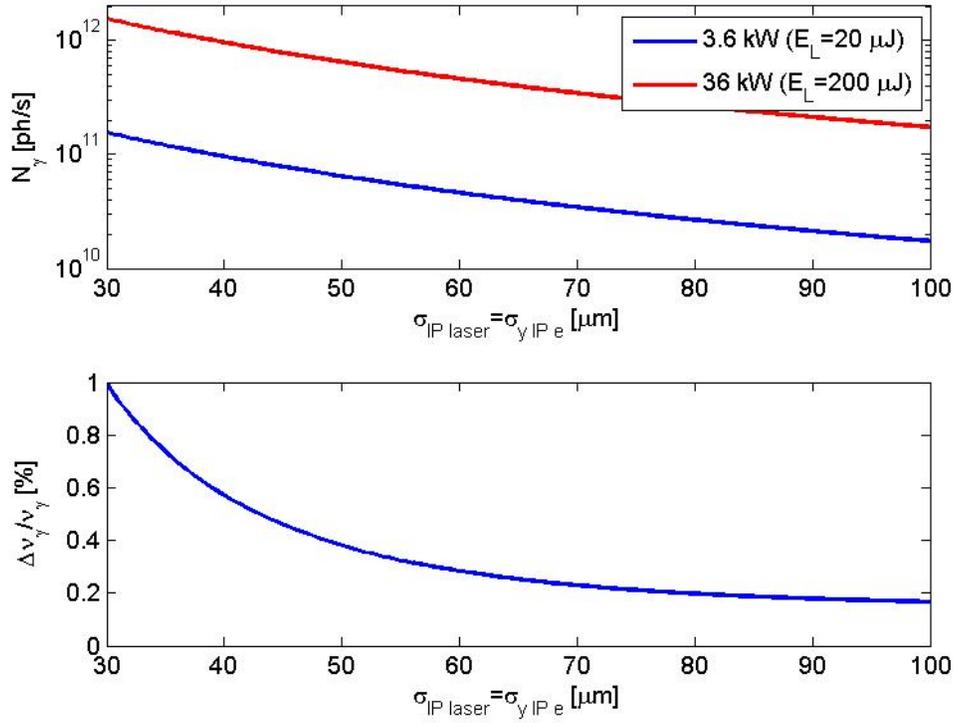

Fig. 3: Total gamma flux and rms bandwidth as a function of the vertical rms laser and e- beam sizes at the IP .

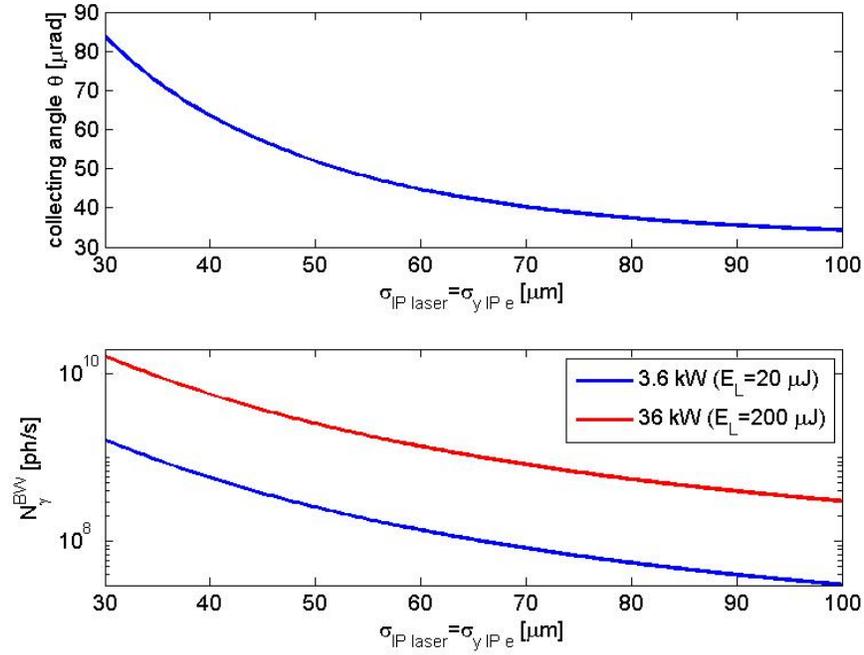

Fig. 4: (top) collecting angle; (bottom) gamma flux within the bandwidth;

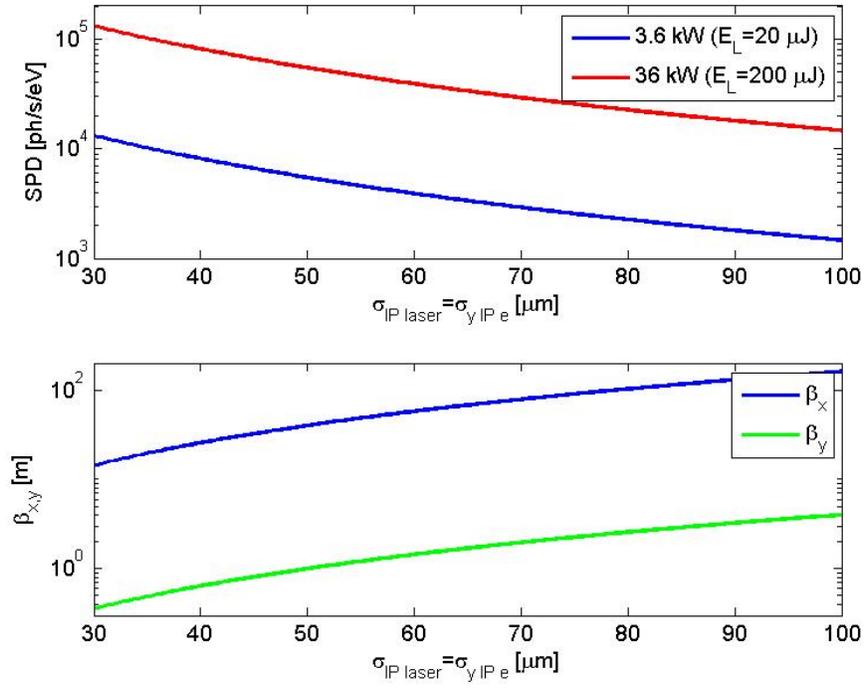

Fig. 5: (top) Spectral density; (bottom) beta functions @ IP.

Table II: DAFNE gamma source parameters

| Parameter | Case 1 | Case 2 |
|---|---|---|
| e⁻ stored current [A] | 1.5 ||
| # of bunches | 60 ||
| bunch charge [pC] | 8152 ||
| e⁻ beam energy [MeV] | 510 ||
| e⁻ beam energy spread [%] | 0.050 ||
| e⁻ emittance [mm*mrad] | 0.100 ||
| coupling @ IP [%] | 2.5 | 1 |
| e⁻ bunch length [ps] | 50 ||
| e⁻ vertical beam dimension @ IP [μm] | 40.0 | 30.0 |
| e⁻ horizontal beam dimension @ IP [μm] | 1600.0 | 3000.0 |
| e⁻ vertical beta function @ IP [m] | 0.6 | 0.9 |
| e⁻ horizontal beta function @ IP [m] | 25.6 | 90 |
| laser circulating power [kW] | 36.8 ||
| Laser energy per pulse [μJ] | 200.0 ||
| laser wavelength [μm] | 1.0 ||
| Fabri-Perrot resonator length [m] | 1.63 ||
| Laser pulse length [ps] | 20 ||
| laser beam dimension @ IP [μm] | 40 | 30 |
| maximum photon energy [MeV] | 4.94 ||
| Collision angle φ [deg] | 8 ||
| Luminosity [cm⁻²s⁻¹] | $1.4 \cdot 10^{36}$ | $1.1 \cdot 10^{36}$ |
| total gamma flux [photons/s] | $0.96 \cdot 10^{12}$ | $0.75 \cdot 10^{12}$ |
| gamma beam energy spread [%] | 0.57 | 0.21 |
| total gamma flux in the bandwidth [ph/s] | $5.8 \cdot 10^{9}$ | $1.7 \cdot 10^{9}$ |
| spectral density SPD [ph/s/eV] | 81533 | 64158 |
| maximum collecting angle [μrad] | 63 | 39 |

## 4       Perturbation in DAFNE beam dynamics

The electrons colliding with photons emit gamma rays with energies proportional to the square of the electron energy (see eq. (2)) and with a typical average spectrum (as reported in the following). This causes a radiation damping but also a quantum excitation of the beam energy since the gammas are stochastically emitted in quantum. Moreover, the gamma rays are emitted at different angles and there is a typical correlation between the energy and the angle as given in eq. (2). The emission of gamma rays in different directions with different energies causes quantum excitation but, on the other hand, it also introduces a radiation damping. The final transverse emittance and the energy spread are due to the balance between these effects and those due to the synchrotron light emission in the ring. Looking at the machine parameters given in Table II it is straightforward to recognize that the average interval between two consecutive collisions of an electron with a photon is much longer than the damping time of the machine and, as a consequence, the beam dynamics is completely dominated by the dynamics of the electron ring without the laser.

There is also another important effect to be considered. If the emitted gamma ray causes an energy variation of the electron above the energy acceptance of the machine, the electron is lost. Therefore the DAFNE energy acceptance has to be larger than the maximum gamma ray energy to avoid a strong reduction of the beam lifetime. If the energy acceptance is dominated by the RF acceptance (as expected) this last condition imposes a constraint to the RF voltage that has to satisfy the following equation:

$$\left.\frac{\Delta E}{E}\right|_{max} = \sqrt{\frac{2\hat{V}_{RF}(\sin\phi_s - \phi_s \cos\phi_s)}{\pi h_N \alpha_c E/e}} > \frac{4ch\gamma^2}{\lambda_L E} \Rightarrow \hat{V}_{RF} > \frac{8c^2 h^2 \gamma^3 \pi h_N \alpha_c}{\lambda_L^2 (\sin\phi_s - \phi_s \cos\phi_s)(E_0/e)e^2}$$

(5)

where $\Delta E/E|_{max}$ is the energy acceptance, $\phi_s$ is the synchronous phase, $\alpha_c$ is the momentum compaction and the other quantities are defined in Table I. Assuming the typical DAFNE ring parameters (see Table I and $\alpha_c$=0.02, $\phi_s \sim \pi/2$) and gamma energies <5 MeV the previous equation imposes $V_{RF}$>180 kV that is compatible with the present DAFNE RF system. From the same equation it is also clear that, if we want to operate at smaller wavelength of the collision laser, we have to increase the RF voltage and/or reduce the momentum compaction with respect to the present typical values. In the eighth section we discuss the case of a different wavelength and the case of different DAFNE energies.

To evaluate the effect of the collisions on the longitudinal and transverse beam dynamics there are two possible different approaches: a theoretical one and particle tracking. The second approach will be discussed in the next section. A simple theoretical model for round beams at the IP is given in the paper by Wang and Ruth [24]. In this model the laser field is considered like a static wiggler. The average damping rate to the normalized transverse emittances can be calculated by the following formula:

$$\Gamma_{x,y}^{laser} = \frac{1}{\tau_{x,y}^{laser}} = \frac{\Delta E_\gamma / E}{T_{rev}}$$

(6)

Where $\Delta E_\gamma$ is the average energy loss of one e$^-$ after passing through the laser and $T_{rev}$ is the revolution time.

The average quantum excitation of the normalized transverse emittances is:

$$\left\langle \frac{d\varepsilon_{x,y}^n}{dt} \right\rangle_{QE\_laser} = \frac{3}{10} \frac{\lambda_c}{\lambda_L} \frac{\Delta E_\gamma}{E} \frac{\beta_{x,y}^*}{T_{rev}}$$

(7)

Where $\lambda_c$ is the Compton wavelength of the electron (2.43·10$^{-12}$ m) and $\beta_{x,y}^*$ are the electron beta functions at IP.

In the longitudinal plane the average damping rate of the energy spread is given by:

$$\Gamma_{\sigma_E}^{laser} = \frac{1}{\tau_\sigma^{laser}} = 2\frac{\Delta E_\gamma / E}{T_{rev}}$$

(8)

While the average quantum excitation to the energy spread can be calculated by:

$$\left\langle \frac{d(\sigma_E^2)}{dt} \right\rangle_{QE\_laser} = \frac{7}{10} \frac{\hbar \omega_m \Delta E_\gamma}{T_{rev}}$$

(9)

Where $\omega_m$ is the maximum angular frequency of the emitted gamma rays ($\omega_m = 4\gamma^2 \omega_L = 8\pi\gamma^2 c/\lambda_L$).

On the other hand the damping rates and quantum excitations due to the synchrotron emission in the ring are given by:

$$\begin{cases} \left\langle \frac{d\varepsilon_{x,y}^n}{dt} \right\rangle_{QE\_ring} = \frac{\varepsilon_{x,y}^n}{\tau_{x,y}^{ring}}; & \Gamma_{x,y}^{ring} = \frac{1}{\tau_{x,y}^{ring}}, \tau_{x,y}^{ring} \cong \frac{\tau_{damp}}{4} = \text{damping time of the transverse emittance} \\ \left\langle \frac{d\sigma_E^2}{dt} \right\rangle_{QE\_ring} = \frac{\sigma_E^2}{\tau_{\sigma_E}^{ring}}; & \Gamma_{\sigma_E}^{ring} = \frac{1}{\tau_{\sigma_E}^{ring}}, \tau_{\sigma_E}^{ring} = \frac{\tau_{damp}}{2} = \text{damping time of the energy spread} \end{cases}$$

(10)

Where $\sigma_E$ and $\tau_{damp}$ are the equilibrium energy spread and the damping time of the DAFNE ring. The final equilibrium emittances and energy spread can be calculated by adding the two contributions in the transverse and longitudinal planes:

$$\begin{cases} \tau_{x,y}^{TOT} = \left(1/\tau_{x,y}^{ring} + 1/\tau_{x,y}^{laser}\right)^{-1} \\ \left\langle \dfrac{d\varepsilon_{x,y}^n}{dt} \right\rangle_{QE\_TOT} = \left\langle \dfrac{d\varepsilon_{x,y}^n}{dt} \right\rangle_{QE\_laser} + \left\langle \dfrac{d\varepsilon_{x,y}^n}{dt} \right\rangle_{QE\_ring} \\ \left.\varepsilon_{x,y}^n\right|_{TOT} = \tau_{x,y}^{TOT} \left\langle \dfrac{d\varepsilon_{x,y}^n}{dt} \right\rangle_{QE\_TOT} \end{cases}$$

$$\begin{cases} \tau_{\sigma_E}^{TOT} = \left(1/\tau_{\sigma_E}^{ring} + 1/\tau_{\sigma_E}^{laser}\right)^{-1} \\ \left\langle \dfrac{d\sigma_E^2}{dt} \right\rangle_{QE\_TOT} = \left\langle \dfrac{d\sigma_E^2}{dt} \right\rangle_{QE\_laser} + \left\langle \dfrac{d\sigma_E^2}{dt} \right\rangle_{QE\_ring} \\ \left.\sigma_E^2\right|_{TOT} = \tau_{\sigma_E}^{TOT} \left\langle \dfrac{d\sigma_E^2}{dt} \right\rangle_{QE\_TOT} \end{cases}$$

(11)

If we want to apply the previous theoretical model to our case (flat e- beam vs. round laser beam) one has to evaluate the average energy loss $\Delta E_\gamma$. This quantity can be evaluated starting from the formula that gives the average number of scattered photons into a given bandwidth for a single electron in a single turn [24]:

$$\frac{d\overline{N}_\gamma}{d\omega} = \frac{3\Delta E_\gamma}{\hbar \omega_m^2}\left[1 - 2\frac{\omega}{\omega_m} + 2\left(\frac{\omega}{\omega_m}\right)^2\right]$$

(12)

From this formula we obtain:

$$\overline{N}_\gamma = \int_0^{\omega_m} \frac{d\overline{N}_\gamma}{d\omega} d\omega = \frac{2\Delta E_\gamma}{\hbar \omega_m} \Rightarrow \Delta E_\gamma = \frac{\overline{N}_\gamma \hbar \omega_m}{2} = \frac{N_\gamma T_{rev} \hbar \omega_m}{2 N_e N_{ph}}$$

(13)

Where $N\gamma$ is given by eq. (1).

The previous formulas, applied to the DAFNE case (assuming the parameters of Table I), give the results plotted in Fig. 6. The plots show a negligible effect on the DAFNE beam dynamics due to the interaction with the laser.

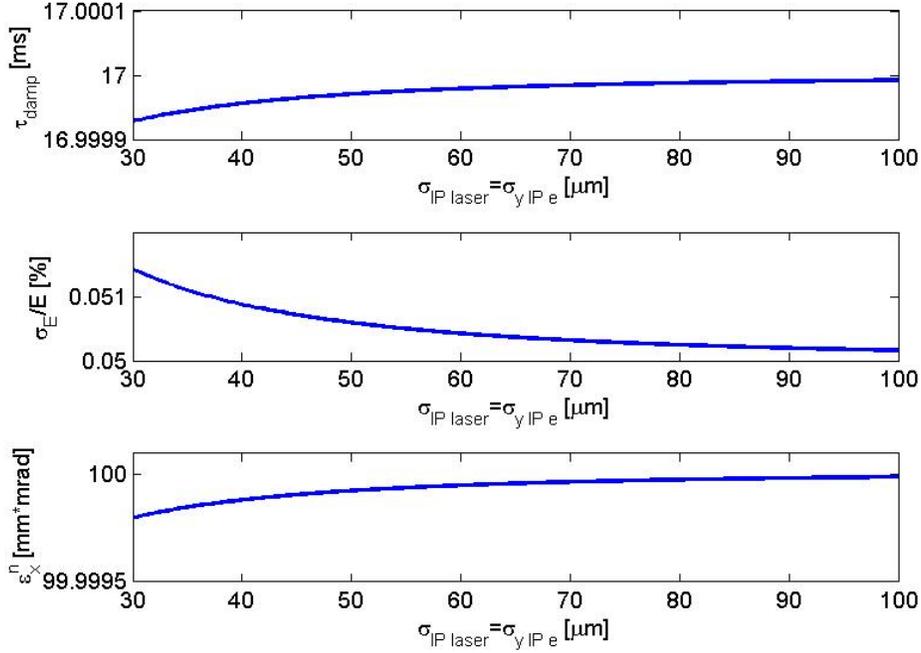

Fig. 6: theoretical damping time, energy spread and normalized emittance of electron beam considering the interaction with the laser.

### 4.1 Maximum achievable energy of the gamma beam

According to eq. (1) the energy of the gamma beam can be changed by varying the energy of the collision laser or the energy of the electron beam. As already pointed out, to avoid a strong reduction of the beam lifetime, it is important to maintain the maximum energy of the gamma photons within the energy acceptance of the ring. The maximum energy of the gamma beam as a function of the electron beam energy for $\lambda_L$ equal to 10, 1 and 0.5 μm are given in Fig. 7. In the same figure it is also reported the required energy acceptance in percent. Energy up to 700 MeV can be, in principle, reached in the DAFNE electron ring by increasing the magnetic elements strength. Typical DAFNE energy acceptances are <1.5% (unless considering very high RF voltages and/or very low momentum compaction factors). As clear from Fig. 7, this fact forces the wavelength of the collision laser to $\lambda_L \geq 1 \mu m$. The resulting energy of the gamma photons can be then tuned reasonably in the range 2.5-9 MeV if we have $\lambda_L = 1 \mu m$ and 250-900 keV if we have, as example, $\lambda_L = 10 \mu m$. In this second case, nevertheless, because of the available technology, lower laser power will be available and, as a consequence, reduced gamma fluxes can be achieved.

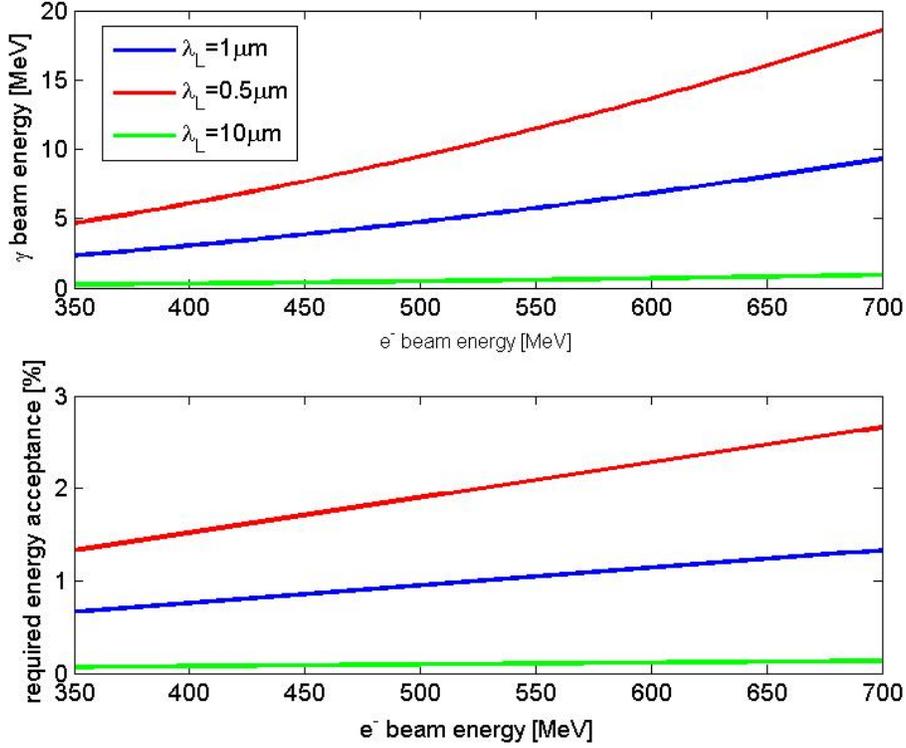

Fig. 7: (top) Maximum energy of the gamma beam as a function of the electron beam energy for $\lambda_L$ =10-1-0.5µm. (bottom) Required energy acceptance.

## 5  CAIN simulation results

In order to validate all theoretical predictions, simulations with the tracking code CAIN [25] have been performed. The results are given in Fig. 8-10 for the parameters reported in Table II-case 1. From these simulations the total gamma flux is $0.88 \cdot 10^{12}$ photons/s, which, within few percent, corresponds to the results obtained by theoretical calculation as given in Table II. The gamma beam energy spread is 0.83% instead of 0.57% while the total gamma flux in the bandwidth is $5.8 \cdot 10^9$ ph/s that is exactly what given in Table II. The corresponding spectral density is 55993 instead of 81533. We can conclude that all results reported in Table II can be reasonably confirmed by CAIN simulations.

With this tracking code also the longitudinal beam dynamics in the presence of the laser-beam interaction has been simulated. The results confirm that the effect on the electron ring beam dynamics is completely negligible.

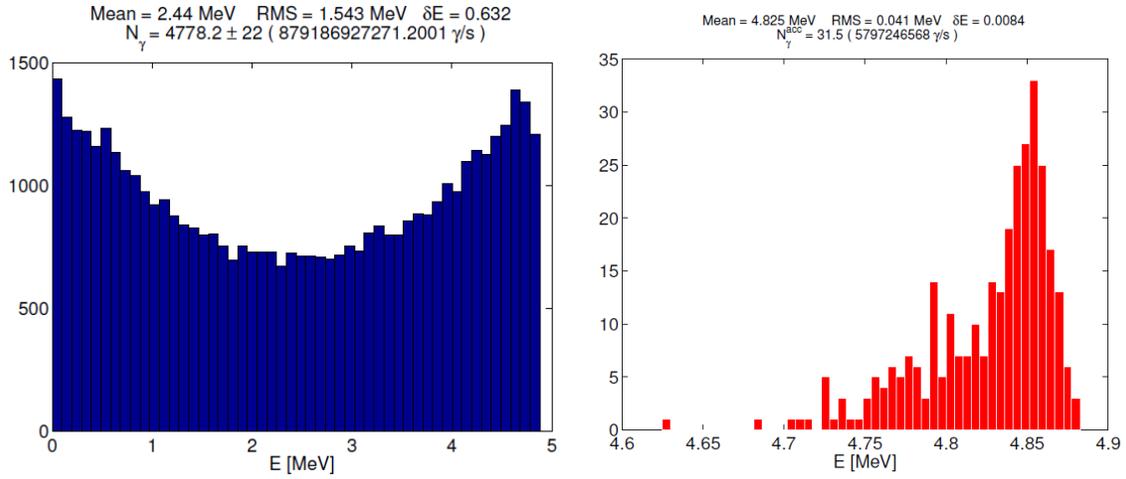

Fig. 8: spectrum with (right) and w/o (left) collimation.

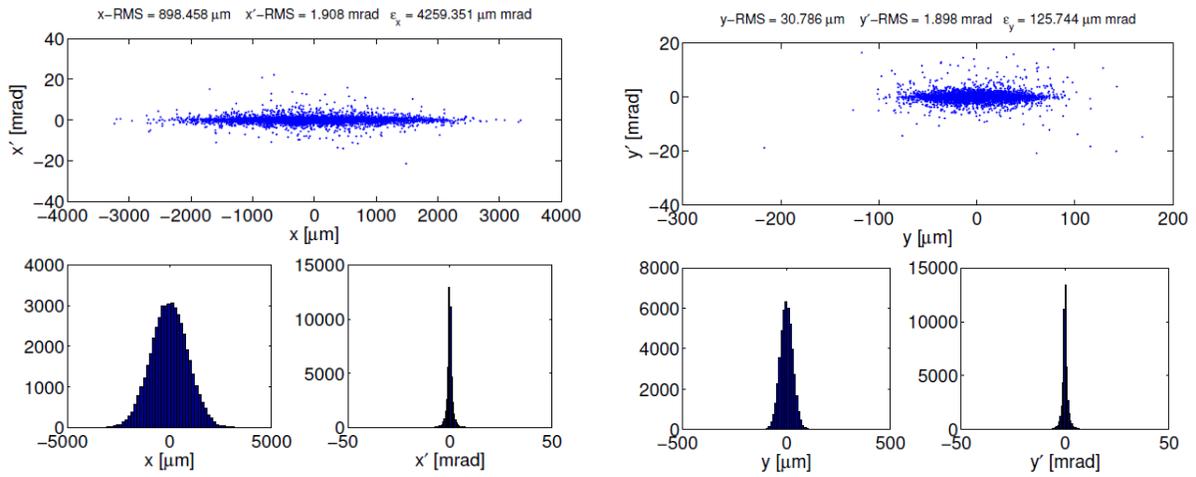

Fig. 9: spatial and angular distribution of the emitted radiation w/o collimation.

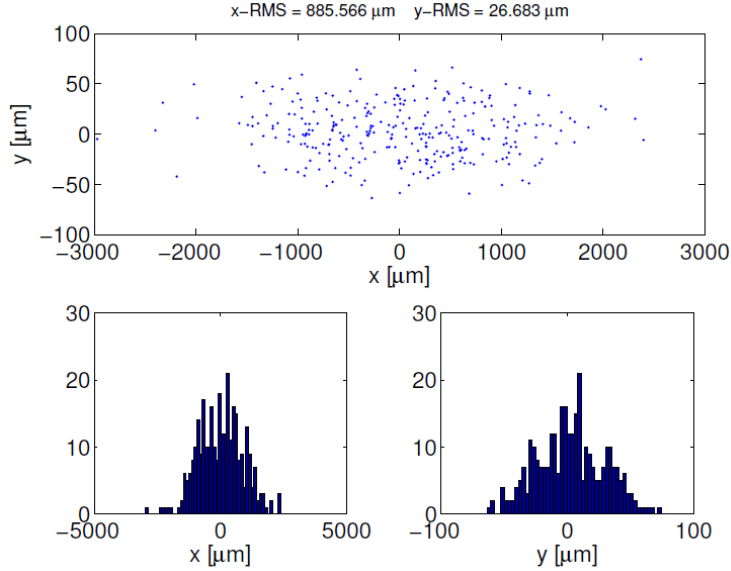

Fig. 10: spatial distribution with collimation.

## 6      Possible implementation

The map of the DAFNE complex is shown in Fig. 11. And the detail of a possible implementation is given in Fig. 12 with the experimental hall that can be located near the present Cryogenics hall. The FPC can be placed in the 1st IP region and, in order to direct the gamma rays in the experimental hall, a small angle of about 5 degrees has to be given to the electron beam trajectory. This induces a machine length variation of about 10 cm. The corresponding variation of the RF frequency is ~350 kHz absolutely compatible with the present RF system (actually with this variation we return back to the original DAFNE RF frequency before the implementation of the crab waist collision scheme).

### 6.1      Fabry Perot resonator: possible upgrade of the existing LAL cavity.

As pointed out in the previous sections, the calculations we have done are referred to the case of a FPC with 1 µm laser wavelength, 184 MHz repetition rate and 36 kW of stored power. All these parameters perfectly match with those already obtained in the FPC realized by the LAL Orsay Laboratory and successfully tested in the Mightylaser experiment at ATF [22]. Few details and images of this FPC are given in appendix II. If we want to use this cavity in the DAFNE gamma factory, the cavity repetition rate has to be increased from the current 178.5 MHz to 184 MHz. To do this one needs to shorten the current cavity round-trip length of 1.68 m by 5 cm. This can be done in two ways:

a) the two concave mirrors are kept at the same positions and the two flat mirrors are moved toward the cavity center by 5/4cm=1.25cm;
b) The four mirrors are moved toward the cavity center by 5/8cm=0.6cm. This would require using another radius of curvature for the concave mirrors.

There are enough margins to make these modifications.
As done in ATF a dedicated shielded chamber has to be inserted in the FPC to reduce the beam coupling impedance and, therefore, the problems related to beam-cavity interaction.

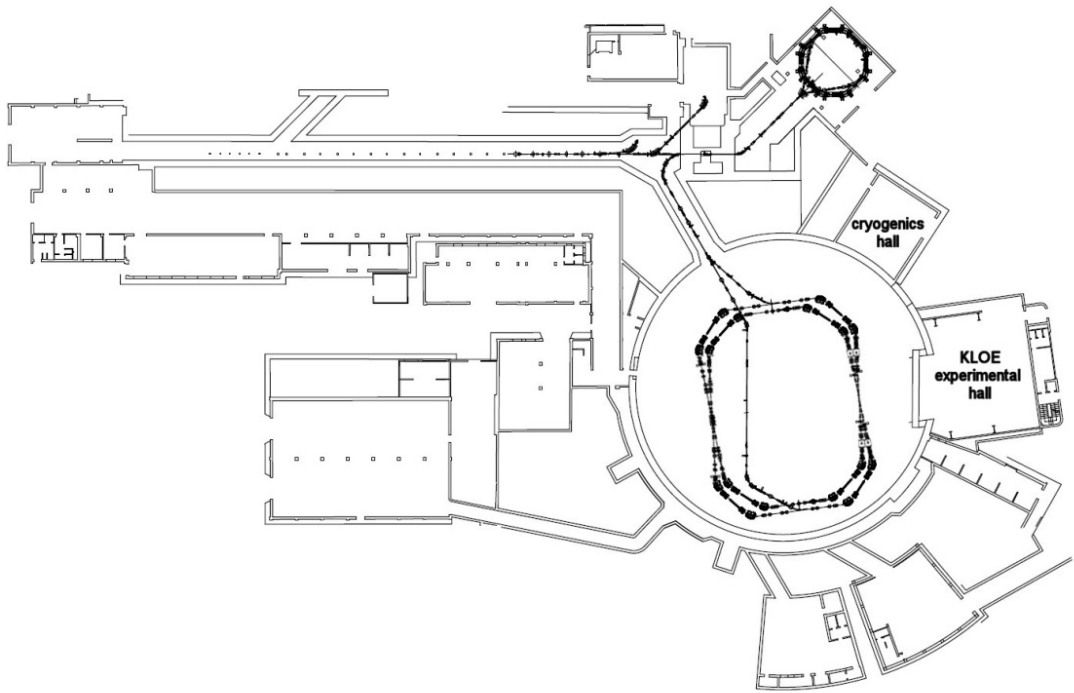

Fig. 11: map of the DAFNE complex.

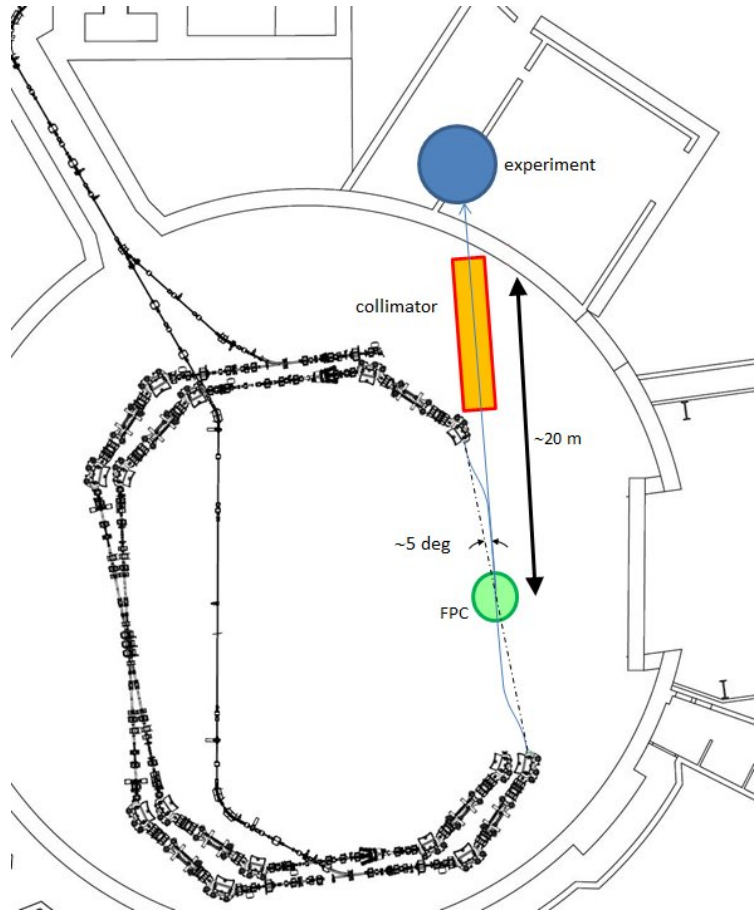

Fig. 12: possible implementation of the gamma beam factory.

## 7  Preliminary studies on low emittance optics

The DAFNE lattice is flexible and allows varying the emittance in a large range. A preliminary version of a low emittance lattice with a horizontal natural emittance of 0.045 mm mrad (a factor ~6 lower than the present DAFNE lattice used for the collider operation) has been studied and is presented in this section. This lattice has been used to study the effects that influence the beam dimensions at the interaction point, such as Intra Beam Scattering (IBS) and bunch lengthening. These effects produce an increase of the longitudinal and transverse emittance (with respect to the natural one) as a function of the bunch current. The conclusion of this study is that the beam parameters listed in Table II can be achieved with a certain flexibility. The next step is the design of a lattice optimized to satisfy the parameters listed in Table II, with a layout of the straight section optimized for the installation of the FPC.

The optical functions in half of the ring are shown in Fig. 13, the other half ring is symmetric. For this study the ring has been assumed symmetric, with the two straight sections for the interaction regions both equal to the present IR2. The reduction of the emittance is obtained by lowering the dispersion function in the wiggler magnets. As a result the dispersion in the interaction regions is increased. The lattice parameters are given in the first column of Table III.

The IBS emittance growth is calculated using a model based on the K. Bane's high energy approximation [26, 27]. The average growth rates are found from the growth rates at each point in the lattice, by integrating over the circumference. The lattice natural emittances are assumed as equilibrium values at low bunch current and the equilibrium emittances at high current, in the presence of radiation and IBS, are found by iteration. The results for the nominal bunch charge of 8.15 nC are listed in Table III, second column. With these parameters the beam sizes growth due to the IBS is a factor 1.9 for the transverse dimensions and a factor 1.2 for the longitudinal ones. Since the natural emittance of the lattice is half of the value given in Table II, also taking into account the IBS effect, the beam sizes at the interaction point are, within 10%, equal to those listed in Table II.

Another possible source of beam sizes increase at the IP is the bunch lengthening effect. The DAFNE bunch length as a function of current has been measured for the Siddharta configuration [28]. The electron ring measurements are well fitted by using the bunch lengthening formula given in [30] with an impedance value Z/n equal to 0.45 Ohm. For the present DAFNE configuration, dedicated to the KLOE2 experiment, the impedance has been further reduced by various vacuum chamber modifications and a value of 0.3 Ohm has been reached [29].

In order to obtain a quick but rather rough estimate of the expected bunch length above the microwave instability threshold the following scaling formula [30], which fits rather well the measured data in ref. [28] and [29], has been used with Z/n=0.3 Ohm.

$$\left(\frac{\sigma_z}{R}\right) \approx \left(\frac{2}{\pi}\right)^{1/6} \left(\frac{2\pi I_b}{h_N e V_{RF} \sin\phi_s}\right)^{1/3} \left(\frac{Z}{n}\right)^{1/3}$$

(14)

where R is the average ring radius, $I_b$ is the bunch current and the other quantities have been defined in the previous equations. For the parameters listed in Table III a bunch lengthening of a factor ~2.8 is estimated with a corresponding reduction of the bunch density. As a consequence the transverse emittance growth due to IBS is reduced. A drawback is the associated increase of the energy spread that, in the presence of a high dispersion at the IP, would give an increase of the transverse beam size. Bunch lengthening simulations performed for the original DAΦNE vacuum chamber [31] have shown that the energy spread increase is smaller than the bunch lengthening. These simulations can be repeated taking into account the wake potential of the present vacuum chamber in order to get a more precise estimate of the expected energy spread and correctly evaluate the IBS effect.

The next step in the lattice design is the design of a lattice satisfying the parameters listed in Table II when taking into account both bunch lengthening and IBS effect. In particular the natural emittance will be increased up to ~0.08 mm mrad by increasing the dispersion function in the wigglers and, correspondingly, reducing it at the IP. This lattice will have smaller beam sizes at the IP for the same energy spread. Moreover the layout of the straight section will be optimized for the installation of the FPC.

Table III: preliminary optics parameters of a low emittance DAFNE lattice

|  | W/O IBS | W IBS |
|---|---|---|
| Beam Energy (GeV) | 0.51 | |
| Circumference (m) | 97.53 | |
| $\alpha_c$ | 0.0077 | |
| Horizontal tune $Q_x$ | 4.870 | |
| Vertical tune Qy | 4.075 | |
| Chromaticity x | -6.84 | |
| Chromaticity y | -10.69 | |
| Emittance x (mm mrad) | 4.48E-02 | 8.60E-02 |
| Emittance y (mm mrad) | 1.12E-03 | 2.15E-03 |
| E loss/turn (MeV) | 0.0089 | 0.0089 |
| Transverse damping time (ms) | 37.1 | |
| Relative energy spread rms (%) | 4.00E-02 | 4.72E-02 |
| RF Voltage (kV) | 180 | 180 |
| Energy RF acceptance (%) | 1.5 | 1.5 |
| Bunch length (mm) | 6.65 | 7.9 |
| Touscheck beam lifetime (s) | 533 | 1381 |
| Horizontal beta function @ IP (m) | 18 | 18 |
| Horizontal dispersion @ IP (m) | 2.8 | 2.8 |
| Vertical beta function @ IP (m) | 0.6 | 0.6 |
| Horizontal beam size @ IP (mm) | 1.436 | 1.815 |
| Horizontal beam divergence @ IP (μrad) | 50 | 69 |
| Vertical beam size @ IP (μm) | 26 | 36 |
| Vertical beam divergence @ IP (μrad) | 43 | 60 |

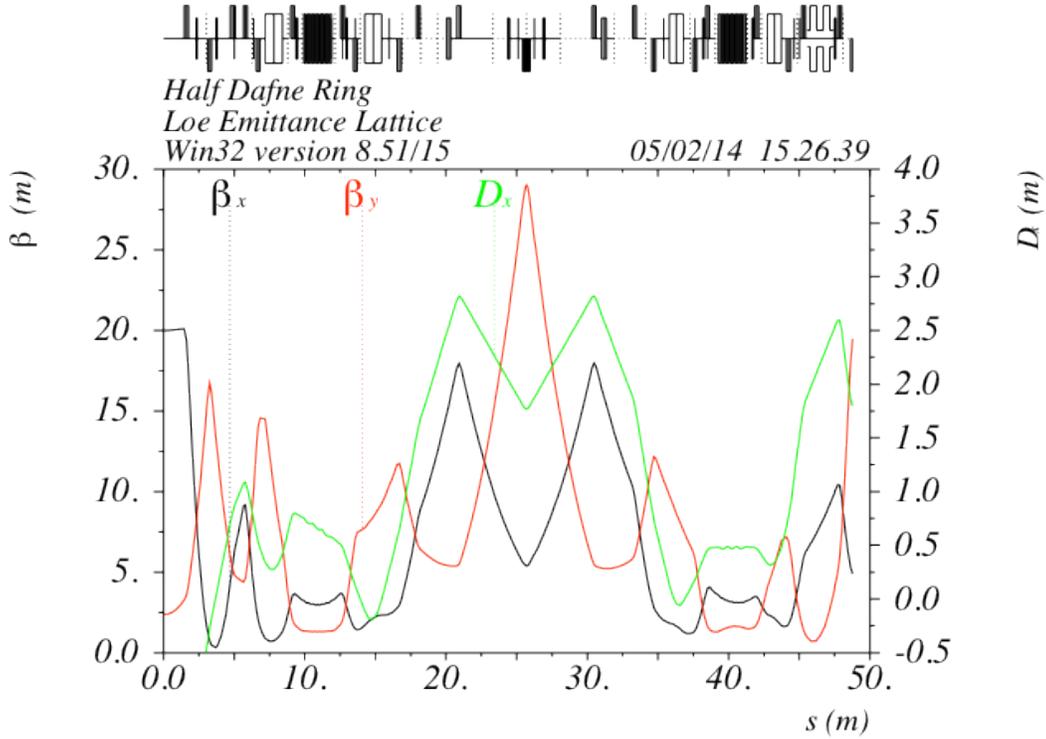

Fig. 13: Low emittance lattice optical functions for half ring starting from the injection straight section. The IR2 straight section is in the middle, the other half ring is mirror symmetric.

## 8    Conclusions

Preliminary studies show that tunable gamma photons can be generated in DAFNE using Compton collisions between the electron beam and a high average power laser pulse amplified in a Fabry-Pérot optical resonator. The very high current storable in the collider rings and the relatively low achievable emittance, allow reaching, in principle, flux and bandwidth comparable (and even better) with the last generation gamma sources. In particular, in the paper, we have analyzed the case of a laser wavelength equal to 1 μm. In this case the photons can be tuned in the 2-9 MeV range and the calculations have shown that the resulting gamma beam source has extremely competitive properties in terms of spectral density ($>5\cdot10^4$ [ph/s/eV]), energy spread (0.5%) and gamma flux ($\sim 10^{12}$ [ph/s]). The main parameters of this new facility have been discussed and the perturbation of the transverse and longitudinal electron beam dynamics has been evaluated. For this experiment there is the possibility to use (with few feasible modifications) the FPC designed and fabricated by the LAL Orsay Laboratory and already tested at ATF. A very preliminary DAΦNE layout to perform experiments with the gamma beam and a first design of a low emittance optics have been presented. Laser wavelength longer than 1 μm can be, in principle, used in the collisions providing gamma rays generation with tunable energies up to 200 keV. In this second case, nevertheless, because of the available technology, lower laser power will be available and, as a consequence, reduced gamma fluxes can be obtained.

**Acknowledgements**

We would like to thank L. Serafini, A. Ghigo and M. Ferrario for their helpful discussions and suggestions.

**Appendix I: Differential Compton cross section for small normalized collecting angles**

The differential Compton cross section in the case of a head on collision between an ultra-relativistic beam and a laser is given in eq. (41) of ref. [16] or in eq. (15) of ref [21]. In both cases assuming $(\gamma\vartheta)^2 \ll 1$ we simply obtain that the cross section for scattering a photon into a cone of angle $\vartheta_{max}$ in laboratory frame is:

$$\sigma(\vartheta_{max}) \cong 4\pi r_0^2 (\gamma\vartheta)^2$$

(A1)

Where $r_0$ is the classical electron radius.

The number of photons per second emitted in this cone is then:

$$N_\gamma^{bw} \cong \sigma(\vartheta_{max}) \cdot L = \sigma(\vartheta_{max}) \frac{N_e N_{ph} f_{coll} \cos(\phi/2)}{2\pi\sqrt{\sigma_{y\_e}^2 + \sigma_{y\_L}^2}\sqrt{(\sigma_{x\_e}^2 + \sigma_{x\_L}^2)\cos^2(\phi/2) + (\sigma_{z\_e}^2 + \sigma_{z\_L}^2)\sin^2(\phi/2)}} \cong$$

$$\cong 6.2 \cdot 10^8 \frac{U_L[J]Q[pC]f_{coll}\cos(\phi/2)}{h\nu_L[eV]\sqrt{\sigma_{y\_e}^2 + \sigma_{y\_L}^2}\sqrt{(\sigma_{x\_e}^2 + \sigma_{x\_L}^2)\cos^2(\phi/2) + (\sigma_{z\_e}^2 + \sigma_{z\_L}^2)\sin^2(\phi/2)}[\mu m^2]}(\gamma\vartheta)^2$$

(A2)

L is the luminosity.

**Appendix II: LAL Fabry Perot resonator used in the Mightylaser experiment at ATF.**

In the framework of the Mightylaser experiment at ATF [22] LAL Orsay designed and built a high finesse optical resonator for high energy gammas production. The cavity is made of four 1 inch diameter high reflective mirrors. Two of them are concave with radii of curvature 500 mm and the two other are flat. The mirror coating were performed by LMA Lyon (absorption loss <1 ppm, diffusion loss <10ppm, transmision ~20ppm). The cavity geometry is non-planar (tetrahedron shape) to provide circularly polarized modes [23]. This cavity is a light circular polarization filter. The four mirrors are mounted inside a frictionless gimbal mount (see Fig. A1). Three stepper motors are used provide one translation and two tilts of the four mirrors. The motors are encapsulated and driven from outside the vacuum vessel. The four mirror mounts are fixed two by two on an invar plate to ensure the stability of the cavity round-trip. The two invar plates are fixed on a stainless steel base plate without constraints. The translations are made using 3 ceramic balls guided in two rails. One of the two flat mirrors is mounted on an annular piezoelectric transducer to lock the cavity round trip frequency to the accelerator frequency. The concave mirror mounts are cut to reduce the laser beam-electron beam crossing angle. The vacuum vessel is a cylinder of 60cm diameter (see Fig. A2).

The obtained performances are listed below:

a) Alignment of the optical axis of the cavity at ~300$\mu$m with respect to the mirror centers.

b) Change the cavity round trip length by translating the 2 flat mirrors to follow the variations of the ATF frequency over 3 years.

c) Vacuum: 3. $10^{-8}$mbar

d) Maximum stored power 36kW for several hours.

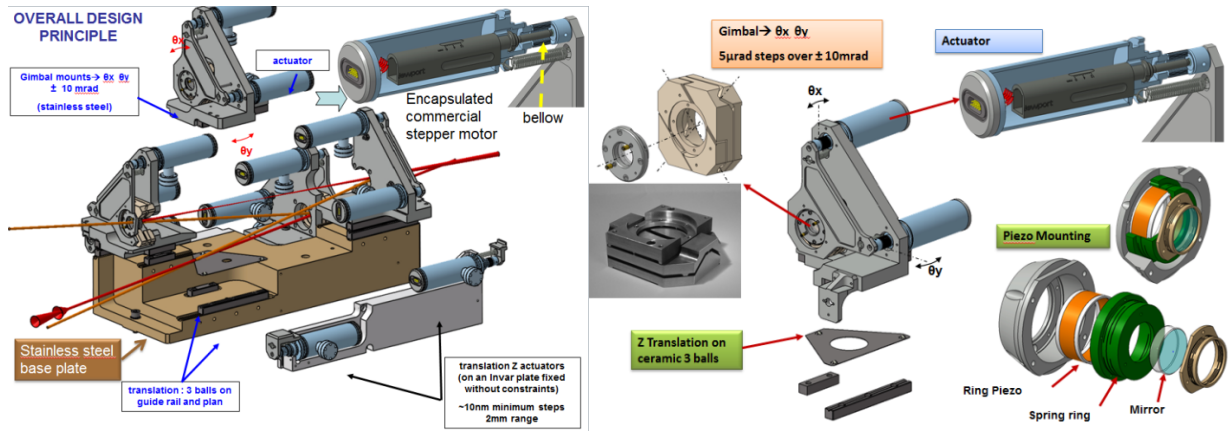

Fig. A1  Schematic view of the cavity mirror mount systems.

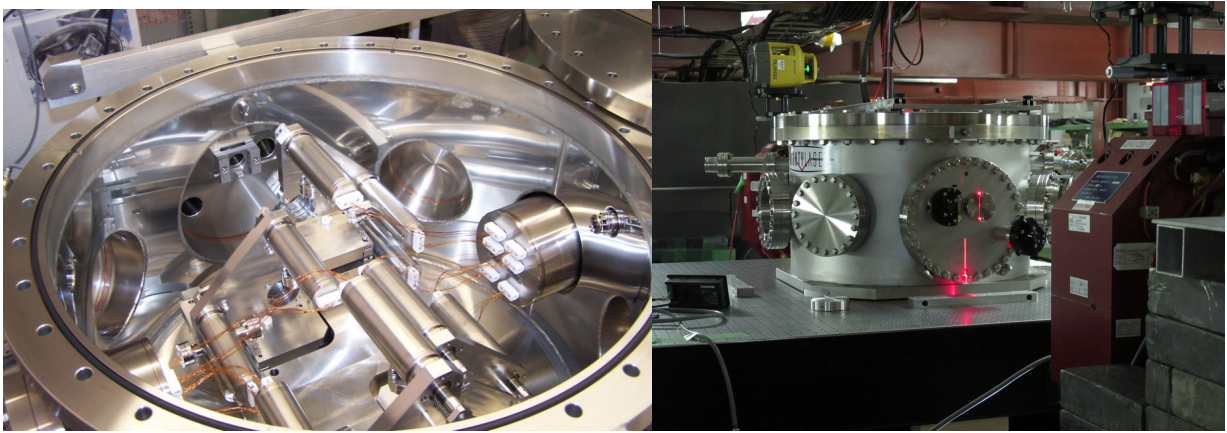

Fig. A2: Pictures of the LAL Orsay FPC.

**References**

[1] Ch. Barty, http://www.eli-np.ro/2011-18_19-aug/gamma-beam-meeting-august-presentation.php and SLACPUB-15149 (2012) available at:

http://www-public.slac.stanford.edu/SciDoc/docMeta.aspx?slacPubNumber=slac-pub-15149

[2] H. Toyokawa, Nucl. Instrum. Methods Phys. Res. A 608, S41–S43, 2009.

[3] Technical Design Report E-Gammas proposal for the ELI-NP Gamma beam System, to be published.

[4] C. Vaccarezza et al., A European Proposal for the Compton Gamma-ray Source of ELI-NP, Proceedings of IPAC2012, New Orleans, Louisiana, USA, 2012.

[5] H. R. Weller and M. W. Ahmed, Mod. Phys. Lett. A 18(23), 1569–1590 (2003); and Y.K. Wu, ACCELERATOR PHYSICS AND LIGHT SOURCE RESEARCH PROGRAM AT DUKE UNIVERSITY, Proceedings of IPAC2013, Shanghai, p. 264, China, 2013.

[6] D. Alesini et al., The Damped C-band RF Structures for the European ELI-NP Proposal, Proceedings of IPAC2013, Shanghai, China, p. 2726, 2013.

[7] Jean-Paul Chambaret1, Răzvan Dabu2, Daniel Ursescu2, The White Book of ELI Nuclear Physics Bucharest-Magurele, Romania, URL www.eli-np.ro/documents/ELI-NP-WhiteBook.pdf.

[8] K. Aoki, et al., Nuclear Instruments and Methods in Physics Research A 516, 228–236, 2004.

[9] K. Kawase, et al., Nuclear Instruments and Methods in Physics Research A 592,154–161, 2008.

[10] V.N. Litvinenko, et al., Phys. Rev. Lett. 78, 4569, 1997.

[11] M. Hosaka, H. Hama, K. Kimura, J. Yamazaki, T. Kinoshita, Nucl. Instr. and Meth. Phys. Res. A 393, 525, 1997.

[12] M.E. Couprie, et al., J. Phys. B: At. Mol. Opt. Phys. 32, 5657, 1999.

[13] A. Bacci, D. Alesini, P. Antici, M. Bellaveglia, R. Boni et al., Electron Linac design to drive bright Compton back-scattering gamma-ray Sources, J. Appl. Phys. 113, 194508, 2013;

[14] L. Serafini et al., TUPO008, Proceedings of IPAC2011, p. 1461, San Sebastián, Spain 2011.

[15] V. Petrillo et al., Nucl. Instrum. Methods Phys. Res. A 693, 109–116, 2012.

[16] W. Brown and F. Hartemann, Phys. Rev. ST– Accel. Beams 7, 060703, 2004.

[17] P. Tomassini et al., Appl. Phys. B: Lasers Opt. 80, 419–436, 2005.

[18] G.Vignola et al., Proceedings of PAC93, p.1993, 1993.

[19] C. Milardi et al., Proceedings of IPAC2011, THPZ004, San Sebastián, Spain, 2011.

[20] M. Zobov et al., Phys. Rev. Lett. 104, 174801, 2010.

[21] J. Stepanek, Nuclear Instruments and Methods in Physics Research A 412, 174-182, 1998.

[22] T Akagi et al, Production of gamma rays by pulsed laser beam Compton scattering off GeV-electrons using a non-planar four-mirror optical cavity, JINST 7 P01021, 2012

[23] J Bonis et al, Non-planar four-mirror optical cavity for high intensity gamma ray flux production by pulsed laser beam Compton scattering off GeV-electrons, JINST 7 P01017, 2012.

[24] Zhirong Huang and Ronald D. Ruth, Laser-Electron Storage Ring, Phys. Rev. Lett. 80, 976, 1998.

[25] http://lcdev.kek.jp/~yokoya/CAIN/Cain242/CainMan242.pdf

[26] K. Kubo, K. Mtingwa, A. Wolski, "Intrabeam scattering formulas for high energy beams", Phys. Rev. STAB 8, 081001, 2005.

[27] K. Bane, in Proceedings of the 8th European Particle Accelerator Conference, Paris, France, 2002 EPS-IGA and CERN, Geneva, p. 1443, 2002.

[28] F. Marcellini, D. Alesini, P. Raimondi, G. Sensolini, B. Spataro, A. Stella, S. Tomassini and M. Zobov, COUPLING IMPEDANCE OF DAΦNE UPGRADED VACUUM CHAMBER, TUPP051, Proceedings of EPAC08, Genoa, Italy, 2008

[29] C. Milardi, et al., DAΦNE TUNE-UP FOR THE KLOE-2 EXPERIMENT, THPZ004, Proceedings of IPAC2011, San Sebastián, Spain, 2011

[30] A.W.Chao, J.Gareyte, Part.Accel.25:229, 1990.

[31] M.Zobov et al., e-Print Arxiv: physics/0312072, DAFNE Technical Note BM-3, 1998.

[32] T. Suzuki, General formulae of luminosity for various types of colliding beam machines, KEK note 76-3, KEK, Tsukuba Japan, 1976.